\documentclass[a4paper,UKenglish,cleveref, autoref, thm-restate]{lipics-v2019}
\usepackage{tgtermes}
\usepackage[T1]{fontenc}
\usepackage{graphicx}
\usepackage{amsfonts,amsmath,amsthm,amssymb,dsfont}
\usepackage{xfrac,nicefrac}
\usepackage{mathdots}
\usepackage{bm,bbm}
\usepackage{url}
\usepackage{paralist}
\usepackage{enumerate}
\usepackage[normalem]{ulem}
\usepackage{xspace}
\xspaceaddexceptions{]\}}
\usepackage{cleveref}
\usepackage{comment}
\usepackage{tabu}
\usepackage{framed}
\usepackage{float,wrapfig}
\usepackage{tikz}
\usepackage[usenames,dvipsnames]{pstricks} 

\usepackage{enumitem}
\usepackage{adjustbox}
\usepackage{algorithm}
\usepackage{algpseudocode}

\algnewcommand{\algorithmicand}{\textbf{AND }}
\algnewcommand{\algorithmicor}{\textbf{OR }}
\algnewcommand{\algorithmicxor}{\textbf{XOR }}
\algnewcommand{\algorithmicnot}{\textbf{NOT }}
\algnewcommand{\OR}{\algorithmicor}
\algnewcommand{\AND}{\algorithmicand}
\algnewcommand{\XOR}{\algorithmicxor}
\algnewcommand{\NOT}{\algorithmicnot}
\algnewcommand{\var}{\texttt}
\algnewcommand{\algorithmicbreak}{\textbf{break}}
\algnewcommand{\Break}{\State \algorithmicbreak}

\usetikzlibrary{hobby}
\usetikzlibrary{decorations.markings}
\usetikzlibrary{quotes,angles}

\theoremstyle{plain}

\newtheorem*{prob*}{Problem}

\theoremstyle{definition}

\def\ShowAuthNotes{1}
\ifnum\ShowAuthNotes=1
\newcommand{\authnote}[2]{\ \\ \textcolor{red}{\parbox{0.9\linewidth}{[{\footnotesize {\bf #1:} { {#2}}}]}}\newline}
\else
\newcommand{\authnote}[2]{}
\fi

\renewcommand{\epsilon}{\varepsilon}

\title{A natural extension to the convex hull problem and a novel solution}

\author{Xiao Mao}{Massachusetts Institute of Technology}{xiao\_mao@mit.edu}{https://orcid.org/0000-0002-1224-9730}{}

\authorrunning{X. Mao} %TODO mandatory. First: Use abbreviated first/middle names. Second (only in severe cases): Use first author plus 'et al.'

\Copyright{Xiao Mao} %TODO mandatory, please use full first names. LIPIcs license is "CC-BY";  http://creativecommons.org/licenses/by/3.0/


 %TODO mandatory; please add comma-separated list of keywords

\acknowledgements{I would like to thank Prof. Michael Sipser for advising my research and giving me advice on my writeup.}%optional

\begin{document}

\setcounter{page}{0} \clearpage
\maketitle
\thispagestyle{empty}

\textbf{Author's Note: Unfortunately, the problem studied in this paper was previously covered in \cite{abrahamsen} where a superior algorithm to the one described here was given. The author did not know of the existence of \cite{abrahamsen} at the time this paper was finished.}

\begin{abstract}
We study a natural extension to the well-known convex hull problem by introducing multiplicity: if we are given a set of convex polygons, and we are allowed to partition the set into multiple components and take the convex hull of each individual component, what is the minimum total sum of the perimeters of the convex hulls? We show why this problem is intriguing, and then introduce a novel algorithm with a run-time cubic in the total number of vertices. In the case that the input polygons are disjoint, we show an optimization that achieves a run-time that, in most cases, is cubic in the total number of polygons, within a logarithmic factor.
\end{abstract}

% keywords can be removed
%\keywords{First keyword \and Second keyword \and More}

\tikzset{->-/.style={decoration={
  markings,
  mark=at position .5 with {\arrow{>}}},postaction={decorate}}}

\section{Introduction}

\subsection{Intuition and Significance}

The convex hull problem is a well-studied problem in computational geometry. Given a set of points on a plane, the convex hull is the minimum convex set that contains this set of points. Computation of the convex hull has been very extensively-studied. Many algorithms that can compute a convex hull for a set of $n$ points in $O(n \log n)$ time are known \cite{bykat, graham, toussaint}, which is also the lower bound for determining the convex hull of $n$ points \cite{yao}. This decade-old problem has found a wide range of applications in mathematics, statistics, combinatorial optimization, economics, geometric modeling, and ethology, and is one of the most important problems in comptutational geometry.

A well-known property of convex hull is that the boundary of the convex hull is the shortest closed curve that encloses a given set of points. This is a fact that has been written into the famous textbook by De Berg et al. \cite{deberg}. A natural question one can ask is: what will happen if we are allowed to use more than one closed curves, and we want to minimize the sum of their lengths? For a set of discrete points on a plane, the answer will be trivial: one can use an infinitesimal curve in the neighborhood of each point to enclose that point, and thus the answer will be infinitesimal. However, if the input points aren't discrete, the answer is no longer infinitesimal. For example, if the input points are the set of points inside a circle, then any convex curve that encloses the input will have a length that is at least the perimeter of the circle. 

\begin{figure}[ht]

    \caption{The optimal solution is to take the perimeter of the convex hull of the two regions, denoted in red} \label{fig:single}

    \centering

    \begin{adjustbox}{max width=1\textwidth}
    \begin{tikzpicture}
        
        \coordinate (A0) at (0, 0); 
        \coordinate (A1) at (0, 3); 
        \coordinate (A2) at (3, 3);
        \coordinate (A3) at (3, 0); 
        
        \coordinate (B0) at (4, 0.5); 
        \coordinate (B1) at (4, 2.5); 
        \coordinate (B2) at (7, 2.5);
        \coordinate (B3) at (7, 0.5); 
    
        \draw [fill=gray] (0, 0) rectangle (3, 3);
        
        \draw [fill=gray] (4, 0.5) rectangle (7, 2.5);
        
        \draw[red, ultra thick] (A0) -- (A1) -- (A2) -- (B2) -- (B3) -- (A3) -- (A0);

    \end{tikzpicture}
    \end{adjustbox}

\end{figure}
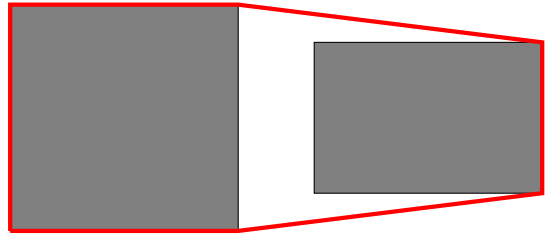

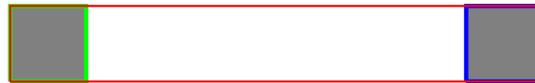
\begin{figure}[ht]

    \caption{The optimal solution is to take the perimeters of the two convex hulls of the two individual regions, denoted in red and blue. The alternative solution of taking the convex hull of the two regions, denoted in red, yields a larger perimeter} \label{fig:double}

    \centering

    \begin{adjustbox}{max width=1\textwidth}
    \begin{tikzpicture}
        
        \coordinate (A0) at (0, 0); 
        \coordinate (A1) at (0, 1); 
        \coordinate (A2) at (1, 1);
        \coordinate (A3) at (1, 0); 
        
        \coordinate (B0) at (6, 0); 
        \coordinate (B1) at (6, 1); 
        \coordinate (B2) at (7, 1);
        \coordinate (B3) at (7, 0); 
    
        \draw [fill=gray] (0, 0) rectangle (1, 1);
        
        \draw [fill=gray] (6, 0) rectangle (7, 1);
        
        \draw[green, ultra thick] (A0) -- (A1) -- (A2) -- (A3) -- (A0);
        
        \draw[blue, ultra thick] (B0) -- (B1) -- (B2) -- (B3) -- (B0);
        
        \draw[red, thick] (A0) -- (A1) -- (A2) -- (B2) -- (B3) -- (A3) -- (A0);

    \end{tikzpicture}
    \end{adjustbox}

\end{figure}

We can see that this model defines an intriguing problem which intuitively is not easy to solve. Suppose we want to enclose two regions $A$ and $B$ on a plane. There are two options: we can either take the convex hull of $A \cup B$, and enclose both sets using the boundary of the convex hull, or we can take individually the convex hull of $A$ and the convex hull of $B$ and use two curves that are their boundaries to enclose these sets. If the two sets are fairly close to each other with regard to their sizes, the optimal solution will be to use one curve, as shown in figure \ref{fig:single}, but if the two sets are fairly far away, the optimal solution will be to use two curves, as shown in figure \ref{fig:double}. 

One can see that, when the input consists of several regions on a plane, we can find a partition of the set of regions, and take the convex hull of each component. The problem becomes to compute the optimal partition. Since the number of way we can partition a set of $n$ regions is exponential in $n$, as the number of regions goes up, the number of options we can have for enclosing the set of regions grows exponentially, making any brute force solutions inefficient. Given that our problem is a pretty straightforward extension of the convex hull problem --- we simply allowed ourselves to use multiple curves, and that an obvious efficient solution does not exist, this new model proves to be intriguing and worth studying.

In this paper, we assume these regions are defined by simple polygons. Moreover, for a non-convex region $A$, any convex set that contains $A$ will contain the convex hull of $A$. Therefore, one can replace each individual input polygon with its convex hull without affecting the answer. Therefore, we will assume that all regions in the input are convex polygons. Our problem can be thereby defined as the following:

\begin{prob*}
    Given a set $C$ of convex polygons, if we partition $C$ into multiple components and take the convex hull of all the convex polygons in each individual component of the partition, what is the minimum total sum of the perimeters of these convex hulls?
\end{prob*}

Note that we do not forbid intersections between the input polygons. Nor do we forbid an input polygon from enclosing another input polygon. The algorithm we will introduce will work even when these special cases are allowed. However, it does seem natural to consider the case where the interiors of all input polygons are disjoint (i.e. no intersection or containment). We will show an improvement to our algorithm in the disjoint case.

A partition is called \textbf{optimal} if it minimizes the answer, namely the minimum total sum of the perimeters of the convex hulls of the components of the partition.

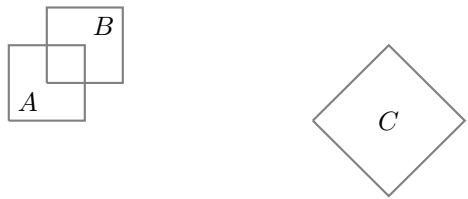
\begin{figure}[ht]

    \caption{Problem example with $C = \{A, B, C\}$} \label{fig:example}

    \centering

    \begin{adjustbox}{max width=1\textwidth}
    \begin{tikzpicture}
        
        \coordinate (A0) at (0, 0); 
        \coordinate (A1) at (0, 1); 
        \coordinate (A2) at (1, 1);
        \coordinate (A3) at (1, 0); 
        
        \coordinate (B0) at (0.5, 0.5); 
        \coordinate (B1) at (0.5, 1.5); 
        \coordinate (B2) at (1.5, 1.5);
        \coordinate (B3) at (1.5, 0.5); 
        
        \coordinate (C0) at (4, 0); 
        \coordinate (C1) at (5, 1); 
        \coordinate (C2) at (6, 0);
        \coordinate (C3) at (5, -1);
        
        \draw[gray, thick] (A0) -- (A1) -- (A2) -- (A3) -- (A0);
        
        \draw[gray, thick] (B0) -- (B1) -- (B2) -- (B3) -- (B0);
        
        \draw[gray, thick] (C0) -- (C1) -- (C2) -- (C3) -- (C0);
        
        \node at (0.25, 0.25) {$A$};
        
        \node at (1.25, 1.25) {$B$};
        
        \node at (5, 0) {$C$};

    \end{tikzpicture}
    \end{adjustbox}

\end{figure}

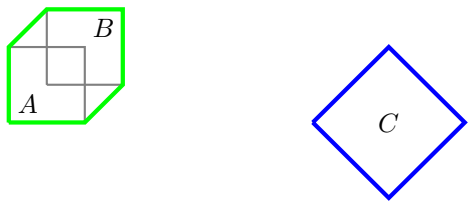
\begin{figure}[ht]

    \caption{Problem example with $C = \{A, B, C\}$ --- Solution} \label{fig:examplesol}

    \centering

    \begin{adjustbox}{max width=1\textwidth}
    \begin{tikzpicture}
        
        \coordinate (A0) at (0, 0); 
        \coordinate (A1) at (0, 1); 
        \coordinate (A2) at (1, 1);
        \coordinate (A3) at (1, 0); 
        
        \coordinate (B0) at (0.5, 0.5); 
        \coordinate (B1) at (0.5, 1.5); 
        \coordinate (B2) at (1.5, 1.5);
        \coordinate (B3) at (1.5, 0.5); 
        
        \coordinate (C0) at (4, 0); 
        \coordinate (C1) at (5, 1); 
        \coordinate (C2) at (6, 0);
        \coordinate (C3) at (5, -1);
        
        \draw[gray, thick] (A0) -- (A1) -- (A2) -- (A3) -- (A0);
        
        \draw[gray, thick] (B0) -- (B1) -- (B2) -- (B3) -- (B0);
        
        \draw[gray, thick] (C0) -- (C1) -- (C2) -- (C3) -- (C0);
        
        \node at (0.25, 0.25) {$A$};
        
        \node at (1.25, 1.25) {$B$};
        
        \node at (5, 0) {$C$};
        
        \draw[green, ultra thick] (A0) -- (A1) -- (B1) -- (B2) -- (B3) -- (A3) -- (A0);
        
        \draw[blue, ultra thick] (C0) -- (C1) -- (C2) -- (C3) -- (C0);

    \end{tikzpicture}
    \end{adjustbox}

\end{figure}

Figure \ref{fig:example} shows a example of the problem with $C = \{A, B, C\}$, and Figure \ref{fig:examplesol} shows the optimal partition $C = \{A, B\} \cup \{C\}$, where the perimeter of the convex hull of $\{A, B\}$ is in green and the perimeter of the convex hull of $\{C\}$ is in blue. The partition is such that the sum of the perimeters of convex hulls of the components is minimum.

\subsection{Relation to Geometric Clustering and Potential Applications} \label{clustering}

A somewhat related problem is to cover points with a fixed number $K$ of closed curves of minimum total perimeter. This problem is an example of the geometric clustering problem, and has been studied extensively. Let $n$ be the number of points. Then the problem can be solved in $O(n ^ 2\log{n})$ time for $K = 2$ \cite{cho} and $O(n ^ {6K - 11})$ time for $K \ge 3$ \cite{capoyleas}. Our model is related to this model in that both models introduce multiplicity to the classic convex hull problem. Unlike the clustering problem, we do not have a constraint on the number of simple closed curves we are allowed to use, and instead we strive for a more general answer: what if we can use any number of simple closed curves. However, the algorithm we are proposing indeed has a weak point: the metric we are using is limited to the sum of perimeters.

The clustering problem has a wide range of applications in pattern recognition, pattern matching, image processing, image understanding and motion analysis \cite{dehne}. With its close relation to the geometry clustering model, our model also applies to some of these applications, and in some cases our models can do more. Cho et al. \cite{cho} gave an example of the application of the geometric clustering model: To detect the existence of a 3-way road intersection, we can track a group of cars. Suppose the cars are initially moving in the same group on a road. An intersection is detected when the cars split into two groups. One can see whether its better cover these cars using only one convex hull or two. If using two convex hulls costs less than using one, then we are confident of the fact that these cars separated into two groups, and can therefore be sure of a 3-way intersection. However, unlike our model, the geometric clustering model only allows for a constant number of convex hulls. By using our model, one can detect when the group of cars split into any arbitrary amount of groups. Not only can our model be used to detect when the cars split into two groups, but also when they split into any arbitrary amount of groups, or merge back into fewer groups. We can not only detect $k$-way intersections with arbitrary $k$, but also detect many other things, enabling us to have a much better understanding of the road information and traffic behavior in an area.

\subsection{Our Results} \label{ourresults}

Although convex hull is a well-known problem and geometry clustering, a closely related problem, has also been extensively studied, the model in our problem appears to have remained untouched. Despite its close relationship with many problems with efficient existing solutions, the solutions to these problems can not be readily used for our problem, and therefore the solution we are proposing is largely original.

Suppose the input consists of $n$ polygons, each polygon has at most $m$ vertices, and the total number of vertices in all polygons is $N$ (therefore $N \le nm$). We will first introduce an $O(N ^ 3)$ time algorithm. Later, we examine the case where the input polygons have disjoint interiors, and we introduce an optimization to the algorithm in this case that achieves an $O(n ^ 3\log{m} + N)$ time, which is much more efficient in the case where the input consists of only a few polygons with a lot of vertices.

Our algorithm is not too complicated. On the most basic level, it simply keeps improving the partition by merging some of its components into one until no improvement can be made by merging. The components we merge in each iteration is computed using a simple dynamic programming process. Both the main process and the dynamic programming sub-process can be described using no more than 20 lines of pseudo-code.

Our algorithm incorporates many existing ideas in computational geometry. The correctness relies on the submodularity of the perimeters of convex hulls. The dynamic programming process uses the well-known ``non-zero rule'' for the point-in-polygon problem, as described in \cite{foley}, which is a variation of the even-odd rule dated back to at least 1962 \cite{shimrat}. The optimization in the disjoint case is straightforward from a well-known fact: the boundary of a convex hull of a set of convex polygons must consist of common outer tangents, and arcs (consecutive segments) of these polygons between two vertices that are the endpoints of common outer tangents --- a fact that has been illustrated in much literature in computational geometry, such as in \cite{nielsen}. When the polygons are disjoint, there will be at most $O(n ^ 2)$ common outer tangents between input polygons, and we thereby bring the number of points of interest from $O(N)$ down to $O(n ^ 2)$. Despite the fact that many of its components are existing ideas, the algorithm itself does not resemble any existing algorithm.

\section{The Algorithm}

\subsection{Intuition}
The main idea of the algorithm comes from the following fact about the submodularity of the perimeter of convex hulls. Let $C$ be a set of convex polygons. Let $\mathcal{H}: 2 ^ C \rightarrow \mathbb{R}$ be a function where for $C ^ {\prime} \subset C$, $\mathcal{H}(C ^ {\prime})$ is equal to the perimeter of the convex hull of $C ^ {\prime}$. Specifically, $\mathcal{H}(\emptyset) = 0$. We have:

\begin{theorem} \label{theo:submodular}
     $\mathcal{H}$ is a submodular function: for $X \subset Y \subset C$, and $x \in C \backslash Y$, $\mathcal{H}(X \cup \{x\}) - H(x) \ge H(Y \cup \{x\}) - H(Y)$.
\end{theorem}

A consequence of submodularity is the following:

If a set $S$ of convex polygons is indivisible, then for all $S ^ {\prime} \supset S$, there exists an optimal partition for $S ^ {\prime}$: $S ^ {\prime} = B_1 \cup B_2 \cup \cdots \cup B_K$ such that $S \subset B_1$.

\begin{theorem} \label{theo:indivisible}
    Let $S {\prime} \subset C$ be such that an optimal way to enclose $S$ is to take the convex hull of the entire set. Namely, for any partition $S = (A_1) \cup (A_2) ^ {\prime} \cup \cdots \cup(A_k)$, $\mathcal{H}(S) \ge \mathcal{H}(A_1) + \mathcal{H}(A_2) + \cdots + \mathcal{H}(A_k)$. Then there exists an optimal partition $C = B_1 \cup B_2 \cdots B_K$, such that $S \subset B_1$. In other words, an optimal way to enclose $C$ is to not split elements in $S$.
\end{theorem}

Both Theorem \ref{theo:submodular} and Theorem \ref{theo:indivisible} will be proven in the appendix.

If a subset $C ^ {\prime} \subset C$ satisfies the condition in Theorem \ref{theo:indivisible}, and \textbf{has a size at least two}, we call this subset \textbf{indivisible}. Due to Theorem \ref{theo:indivisible}, whenever we find a indivisible subset, we can "merge" the polygons in this subset into its convex hull, shrinking the size of the set of polygons without affecting the answer.

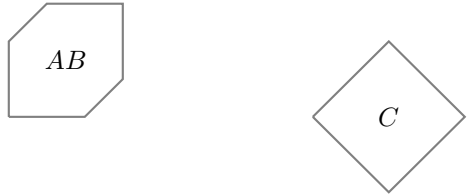
\begin{figure}[ht]

    \caption{Problem Example After Merging} \label{fig:mafter}

    \centering

    \begin{adjustbox}{max width=1\textwidth}
    \begin{tikzpicture}
        
        \coordinate (A0) at (0, 0); 
        \coordinate (A1) at (0, 1); 
        \coordinate (A2) at (1, 1);
        \coordinate (A3) at (1, 0); 
        
        \coordinate (B0) at (0.5, 0.5); 
        \coordinate (B1) at (0.5, 1.5); 
        \coordinate (B2) at (1.5, 1.5);
        \coordinate (B3) at (1.5, 0.5); 
        
        \coordinate (C0) at (4, 0); 
        \coordinate (C1) at (5, 1); 
        \coordinate (C2) at (6, 0);
        \coordinate (C3) at (5, -1);
        
        \draw[gray, thick] (A0) -- (A1) -- (B1) -- (B2) -- (B3) -- (A3) -- (A0);
        
        \draw[gray, thick] (C0) -- (C1) -- (C2) -- (C3) -- (C0);
        
        \node at (0.75, 0.75) {$AB$};
        
        \node at (5, 0) {$C$};

    \end{tikzpicture}
    \end{adjustbox}

\end{figure}

For example, in figure \ref{fig:example} before the set $\{A, B\}$ is a indivisible set, and figure \ref{fig:mafter} shows the set after the merging. We can get optimal answer for the set in figure \ref{fig:mafter} by taking the boundary of each polygon, which is the same as the answer for the original set as shown earlier in figure \ref{fig:examplesol}.

Obviously, a partition is not optimal if and only if there exists a indivisible subset. In that case, the best way to enclose the set of polygons is to take the boundary of each polygon. Therefore, given a set of convex polygons $C$, an algorithm that keeps merging a indivisible set until no such set can be found will yield a set $C ^ {\prime}$ that implies the correct answer for our problem. The optimal partition is implied by what polygons in $C$ each polygon in $C ^ {\prime}$ is merged from. For example, if the input set is $\{A, B, C\}$, and the final set of polygons contains the convex hull of $\{A, B\}$ and $C$. We know the partition is $\{A, B, C\} = \{A, B\} \cup \{C\}$. Since whenever we merge a indivisible subset of size, the number of remaining polygons decreases, merging happens at most $n$ times. Therefore, if we can design an efficient algorithm that finds a indivisible subset, we can solve the problem efficiently.

\subsection{Description of the Algorithm}

A set of polygon is \textbf{optimal} if no indivisible sets exist (i.e. no improvement can be found). For example, the set in figure \ref{fig:mafter} is not optimal, while the set in figure \ref{fig:mafter} is, since merging $AB$ and $C$ increases the answer.

We use the following a way two compare two points in $\mathbb{R} ^ 2$: for two points $u$ and $v$, $u < v$ if either of the following is true:

\begin{itemize}
    \item $u$ has a smaller x-coordinate than $v$.
    \item $u$ has the same x-coordinate as $v$ and $u$ has a smaller y-coordinate than $v$.
\end{itemize}

For a polygon $P$, let $V(P)$ be the set of vertices on $P$. Let $V(C ^ {\prime})$ be the set of vertices in $C ^ {\prime}$. Let $\min V(P)$ be the minimum vertex on $P$ according the order above (i.e. the "lower-left" point if we use a Cartesian plane). If the set of input polygons is $C$, the algorithm first sorts the input polygons $P \in C$ in decreasing order of $\min V(P)$. Then, it maintains an set $C ^ {\prime}$ of polygons that is optimal. For each polygon $P$ in the sorted order, it adds $P$ to $C ^ {prime}$ and keeps $C ^ {\prime}$ optimal by merging. At the end of the process, the answer and the optimal partition will be implied by $C ^ {\prime}$. 

For a subset of $S$ of $C ^ {\prime}$, let $\Delta(S)$ be the change in the sum of perimeters after merging $S$ (i.e. the sum after the merging minus the sum before the merging). Namely,
    
\begin{align*}
    \Delta(U) = H(U) - \sum_{P_U \in U}{H(P_U)}
\end{align*}

In order to make $C ^ {\prime}$ optimal again after adding a new polygon. The algorithm calls a sub-procedure to compute a subset $S$ such that $\Delta(S)$ is minimum (and therefore the improvement we can get by merging $S$ is the greatest), and merges this subset if $\Delta(S) \le 0$. We will prove later with this agenda, we always merge a indivisible set. $C ^ {\prime}$ becomes optimal if $\Delta(S) > 0$, 

The pseudocode is as follows. The sub-procedure FIND\_MIN($C ^ {\prime}$) finds the subset $S$ in $C ^ {\prime}$ such that $\Delta(S)$ is minimum:

\begin{algorithm}
    \caption{}
    \begin{algorithmic}[1]
        \Procedure{MAIN}{$C$}
            \State $C ^ {\prime} \gets \emptyset$
            \For {$P \in C$ in decreasing order of $\min V(P)$}
                \State $C ^ {\prime} \gets C ^ {\prime} \cup P$
                \While {true}
                    \State $S \gets FIND\_MIN(C ^ {\prime})$
                    \If {$\Delta(S) \le 0$}
                        \State In $C ^ {\prime}$, merge $S$ into its convex hull 
                    \Else
                        \Break
                    \EndIf
                \EndWhile
            \EndFor
            \Return $C ^ {\prime}$
        \EndProcedure
    \end{algorithmic}
\end{algorithm}

\newpage

Now we introduce the sub-procedure FIND\_MIN. We compute the subset $S \subset C ^ {\prime}$ with minimum $\Delta(S)$ indirectly: we first find the boundary of the convex hull $H$ (which is a simple closed curve), and then we take the polygons in $C ^ {\prime}$ that it encloses as the subset $S$. For any simple closed curve $X$, define $G(X)$ as the length of $X$ minus the sum of the perimeters of the polygons in $C ^ {\prime}$ that $X$ encloses. By these definitions, $G(H) = \Delta(S)$, and $H$, the boundary of the convex hull of the subset $S$ with minimum $\Delta(S)$, is the simple closed curve such that $G(H)$ is minimum, since simple closed curves that are not boundaries of convex hulls are non-optimal. In other words, if we can find the simple closed curve $h$ that minimizes $G(h)$, then $h = H$. 

Consider $P$, the last polygon added to $C ^ {\prime}$ before calling FIND\_MIN($C ^ {\prime}$). Before $P$ was added to $C ^ {\prime}$, there was no indivisible set in $C ^ {\prime}$. Therefore after adding $P$, a non-divisile set of $C ^ {\prime}$ must contain $P$ or an element that is merged from $P$, and its convex hull must contain $P$. From the order the main procedure adds polygons to $C ^ {\prime}$, if $P$ is the last polygon added, then $\min V(P) = \min V(C ^ {\prime})$. Since $H$ is the boundary of a convex hull of a subset of $C ^ {\prime}$ that contains $P$, $\min V(P)$ must be the minimum point on $H$.

Suppose we traverse $H$ in the clockwise order from $\min V(P)$ until we visit that vertex again, and write down the vertices we visit in order: $H_0 = \min V(P), H_1, H_2 \cdots H_{K - 1}, H_K = \min V(P)$\footnote{we use 0-based indexing.}. For a point $x$, define the angle of $x$, denoted by $\textrm{ANGLE}(x)$, as as the angle between the vector from $\min V(P)$ to $x$ and the positive direction of the y-axis. It is a well-known property that the angles of $H_1, H_2, \cdots H_{K - 1}$ are monotonically increasing \footnote{The definition for convexity is strict: no three vertices on the convex hull can be co-linear.}. Let $p$ be the sequence of length $|V(C ^ {\prime})| + 1$ such that $p_0 = p_{|V(C ^ {\prime})|} = \min V(P)$, and $p_{1 \cdots {|V(C ^ {\prime})|} - 1}$ are the vertices in $V(C ^ {\prime}) \backslash \min V(P)$, sorted in the increasing order of their angles. Then there exists $0 = i_0 < i_1 < \cdots i_{k - 1} < i_K = |V(C ^ {\prime})|$ such that $H_j = i_j$ for all $0 \le i \le K$.

We define an easily computable function $\Gamma: V(C ^ {\prime}) \times V(C ^ {\prime}) \rightarrow \mathbb{R}$ such that $\sum_{0 \le i < K}{\Gamma(H_i, H_{i + 1})} = \Delta(H)$. With this function defined, the problem becomes manageable: we will only need to find indices $0 = i_0 < i_1 < \cdots i_{k - 1} < i_K = |V(C ^ {\prime})|$ such that $\sum_{0 \le i < K}{\Gamma(H_i, H_{i + 1})}$ is minimum, a problem that can be solved with a dynamic programming process: let $f_i$ be the optimal answer for $p_{0 \cdots i}$, then $f_i = \min_{0 \le j < i}{f_j + \Gamma(p_j, p_i)}$. To recover the polygon $H$, we store another array $r$ where $r_i$ records the index $j$ where $f_j + \Gamma(p_j, p_i) = f_i$.

To define the function $\Gamma$, we make use of the following theorem that we will prove later:

\begin{theorem} \label{theo:partial}
    $H$ never "partially encloses" a polygon in $C ^ {\prime}$, that is, for any polygon $x \in C ^ {\prime} \backslash S$, no point inside $x$ (excluding boundary) is enclosed by $H$.
\end{theorem}

From Theorem \ref{theo:partial}, for each polygon $P$ in $C ^ {\prime}$, and any point strictly inside $P$, $H$ encloses $P$ if and only if it encloses that point. According to the non-zero rule, to check whether a point is inside a polygon, we can use the following process: cast a ray from the point in the positive direction of the y-axis, traverse the boundary of the polygon in clock-wise order, and set a counter initially to 0. Whenever an edge on the polygon crosses the ray while going from left to right, we add 1 to the counter, and whenever we cross the ray while going from right to left, we subtract 1 from the counter. The counter will have a value 1 if the point is inside the polygon, and 0 if it is outside the polygon. 

Therefore, for any two distinct vertices $u$ and $v$, we calculate a value $Y$ in this way: set $Y$ initially to the zero, and for each $P \in C ^ {\prime}$, we take a point strictly in $P$, and cast a ray in the positive direction of the y-axis from the point. If by going from $u$ to $v$ we cross the ray while going from left to right, we add the perimeter of $P$ to $Y$, and if by going from $u$ to $v$ we cross the ray while going from right to left, we subtract the perimeter of the polygon from $Y$. We can see that $Y$ is equal to the sum of the perimeters of polygons in $S$. For any curve $l$, $\|l\|$ denotes its length. For two points $a$ and $b$, $a b$ denotes the line segment connecting $a$ and $b$. Then $\Gamma(u, v) = \|u v\| - Y$.

There is still one caveat: we need our final polygon to satisfy the condition required in Theorem \ref{theo:partial}, which can be done by making sure that no line segment involved in the transition during the dynamic programming intersects the interior (excluding boundary) of another polygon. The pseudo-code for the sub-procedure is as follows:

\begin{algorithm}
    \caption{}
    \begin{algorithmic}[1]
        \Procedure{FIND\_MIN}{$C ^ {\prime}$}
            \State $p_0 \gets \min V(P)$
            \State $p_{|V(C ^ {\prime})|} \gets \min V(P)$
            \State $p_{1 \cdots V(C ^ {\prime})} \gets$ vertices in $V(C ^ {\prime}) \backslash \min V(P)$ in increasing order of angles 
            \State $f_0 \gets 0$
            \For {$1 \le i \le |V(C ^ {\prime})|$}
                \State $f_i \gets \infty$
                \For {$0 \le j < i$}
                    \If {the line segment $p_i p_j$ does not intersect the interior (excluding boundary) of any other polygon in $C ^ {\prime}$}
                        \If {$f_j + \Gamma(p_j, p_i) < f_i$}
                            \State $f_i \gets f_j + \Gamma(p_j, p_i)$
                            \State $r_i \gets j$
                        \EndIf
                    \EndIf
                \EndFor
            \EndFor
            \State Recover $H$ using the array $r$.
            \State $S \gets $the set of polygons in $C ^ {\prime}$ that $H$ encloses
            \Return $S$
        \EndProcedure
    \end{algorithmic}
\end{algorithm}

\newpage

To speed up the queries on values of $\Gamma$, we can maintain a lookup table for that function: $g_{i, j}$ stores the value of $\Gamma(p_i, p_j)$. Similarly, we maintain another look-up table $t$ where $t_{i, j}$ is a boolean indicator of whether the line segment $p_i p_j$ is intersection-free. Let $N$ be the total number of vertices in $C$. Whenever a new polygon $P$ of $m$ vertices is added to $C ^ {\prime}$, or removed from $C ^ {\prime}$ due to a merge into a larger convex hull, table $g$ can be updated in $O(1)$ time per entry by testing the intersection between a ray and a segment. 

For the table $t$, to check if a line segment intersects the interior (excluding boundary) of $P$. We check whether the endpoints of the line segment is in $P$, and whether the line segment intersects the boundary of $P$ at non-endpoints. The first can be done by ray-casting and the latter can be done by checking if the line segment intersects some edge on $P$. Update for the table $t$ can be done in $O(m)$ time per entry by going through the edges. Note that whenever we merge a set $S$. There are at most $|S|$ edges on the convex hull of $S$ that are not an edge of a polygon in $S$. Since the edges that are both on the convex hull and some polygon in $S$ will be first deleted and then added back, which cancel each other out, we can save time by ignoring these edges. With this speed up, for each entry, merging the set $S$ takes $O(|S|)$ time. By amortized analysis, the total size of the sets merged, and therefore the number of edges involved, is no more than $2n - 1$. Therefore the total complexity is $O(N + n) = O(N)$ per entry, and the total complexity for all entries is $O(N ^ 3)$. FIND\_MIN will be called $O(n)$ times and dynamic programming takes $O(N ^ 2)$ time per call, yielding an $O(N ^ 2n)$ total time for this part. The total time complexity is $O(N ^ 3 + N ^ 2n) = O(N ^ 3)$.

\subsection{Correctness}

Two questions remained unanswered in the previous section: firstly, we need to show that the subset $S$ of $C ^ {\prime}$ of minimum $\Delta(S)$ is indivisible, if $\Delta(S) \le 0$. Secondly, we want to show that Theorem \ref{theo:partial} is true.

To show $S$ is indivisible, it suffices to prove the following theorem:

\begin{theorem} \label{theo:deltand}
    Given a set $S$ of convex polygons, if for all $S ^ {\prime} \subset S$, $\Delta(S ^ {\prime}) \ge \Delta(S)$, and there exists $P \in S$ such that for all $S ^ {\prime} \subset (S \backslash P)$, $\Delta(S ^ {\prime}) \ge 0$, then $S$ is indivisible.
\end{theorem}

\begin{proof} [Proof of Theorem \ref{theo:deltand}]
    Let $S = A_1 \cup A_2 \cup \cdots A_K$ be a partition for $S$ such that the sum of perimeters of the convex hulls of the components of $A$ is less than the perimeter of the convex hull of $S$. Namely, we have:
    
    \begin{center}
        $\sum_{k}{\mathcal{H}(A_k)} < \mathcal{H}(S)$
    \end{center}
    
    Suppose $A_1$ is the component that contains $P$. Then for all $i > 1$, $\Delta(A_i) \ge 0$. Therefore, $\sum_{k}{\Delta(A_k)} \ge \Delta(A_1)$. Since $A_1 \subset S$, $\Delta(S) \le \Delta(A_1)$. Therefore $\sum_{k}{\Delta(A_k)} \ge \Delta(S)$. Therefore:
    \begin{align*}
        \mathcal{H}(S) - \sum_{k}{\mathcal{H}(A_k)} &= \mathcal{H}(S) - \sum_{P_S \in S}{\mathcal{H}(P_S)} - (\sum_{k}{\mathcal{H}(A_k)} - \sum_{k}{\sum_{P_{A_k} \in A_k}{\mathcal{H}(P_{A_k})}}) \\
                                                                &= \Delta(S) - \sum_{k}{\Delta(A_k)} \\
                                                                &\le \Delta(S) - \Delta(S) \\
                                                                &\le 0
    \end{align*}
    
    This would contradict $\sum_{k}{\mathcal{H}(A_k)} < \mathcal{H}(S)$.
\end{proof}

Finally, to show Theorem \ref{theo:partial}, it suffices to show the following Lemma.

\begin{lemma} \label{lemma:partial}
    If $A$ and $B$ are convex polygons such that there exists point $p$ in the interior of both $A$ and $B$. Then $\mathcal{H}(\{A, B\}) < \mathcal{H}(A) + \mathcal{H}(B)$
\end{lemma}

\begin{proof}
    Consider Theorem \ref{theo:submodular}. We have:
    \begin{align*}
        \mathcal{H}(\{p, B\}) - \mathcal{H}(\{p\})  &\ge \mathcal{H}(\{A, B\}) - \mathcal{H}(\{A\}) \\
        \mathcal{H}(B) - 0                          &\ge \mathcal{H}(\{A, B\}) - \mathcal{H}(\{A\}) \\
        \mathcal{H}(B) + \mathcal{H}(\{A\})         &\ge \mathcal{H}(\{A, B\}))
    \end{align*}
    
    Which proves the lemma.
\end{proof}

\begin{proof} [Proof of Theorem \ref{theo:partial}]
    If $S$ partially contains some $P \in C ^ {\prime} \backslash S$. Then from Lemma \ref{lemma:partial}, $\mathcal{H}(S \cup \{P\}) < \mathcal{H}(S) + \mathcal{H}(P)$. Therefore:
    \begin{align*}
        \mathcal{H}(S \cup \{P\})                                                                     &< \mathcal{H}(S) + \mathcal{H}(P) \\
        \mathcal{H}(S \cup \{P\}) - \sum_{P_S \in S}{\mathcal{H}(P_S)} - \mathcal{H}(P)   &< \mathcal{H}(S) + \mathcal{H}(P) - \sum_{P_S \in S}{\mathcal{H}(P_S)} - \mathcal{H}(P) \\
        \Delta(S \cup \{P\})                                                                                &< \Delta(S)
    \end{align*}
    
    Which contradicts the fact that $\Delta(S)$ is minimum.
\end{proof}

\section{The Disjoint Case and An Optimization}

The algorithm we introduced runs in $O(N ^ 3)$ time, where $N$ is the number of vertices in the input. Consider the case where the number of polygons is constant, while each polygon has a lot of vertices on it. In this case, even a brute force algorithm that checks for each partition of the input polygons will run in $O(N)$ time --- more efficient than our algorithm. The problem for our algorithm is that it does not distinguish a pair of vertices on the same polygon with a pair of vertices that are not on the same polygon. Fortunately, in the most natural case where the the interiors input polygons are disjoint, we can take advantage of this distinction and reach a run time of $O(n ^ 3\log{m} + N)$, where $n$ is the number of polygons and $m$ is the maximum number of vertices on a single polygon.

As mentioned earlier, the boundary of a convex hull of a set of convex polygons must consist of common outer tangents, and arcs of these polygons between two vertices that are the endpoints of common outer tangents. Therefore, we can find the common outer tangents between all pairs of polygons. Since the polygons are disjoint, there are two common outer tangents for each pair. Since there are $O(n ^ 2)$ pair of polygons, there are $O(n ^ 2)$ common outer tangents.

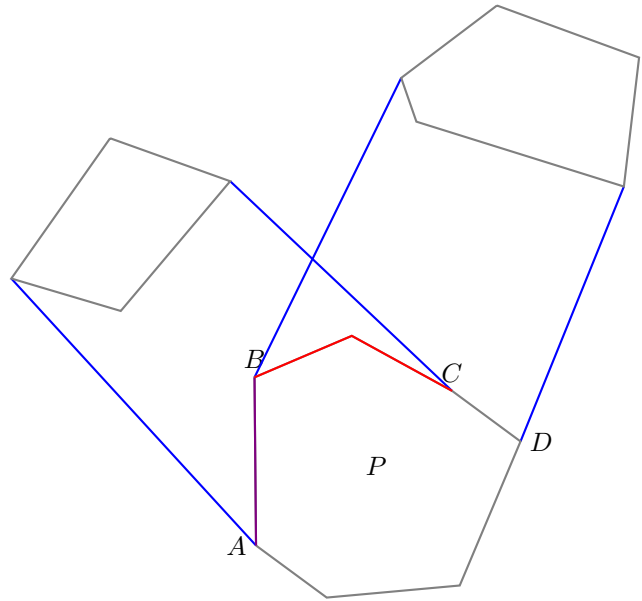
\begin{figure}[ht]

    \caption{Common Outer Tangents} \label{fig:cot}

    \centering

    \begin{adjustbox}{max width=1\textwidth}
    \begin{tikzpicture}
        
        \coordinate[label = left:$A$] (A) at (-1.58, -1.05);
        \coordinate (B) at (-0.65, -1.74); 
        \coordinate (C) at (1.1, -1.58); 
        \coordinate[label = right:$D$] (D) at (1.9, 0.33); 
        \coordinate[label = above:$C$] (E) at (1, 1); 
        \coordinate (F) at (-0.32, 1.73); 
        \coordinate[label = above:$B$] (G) at (-1.6, 1.18); 
        
        \coordinate (H) at (1.59, 6.11); 
        \coordinate (I) at (0.33, 5.15); 
        \coordinate (J) at (0.53, 4.57); 
        \coordinate (K) at (3.26, 3.71); 
        \coordinate (L) at (3.46, 5.42); 
        
        \coordinate (M) at (-3.5, 4.35); 
        \coordinate (N) at (-4.8, 2.49); 
        \coordinate (O) at (-3.36, 2.06); 
        \coordinate (P) at (-1.92, 3.78); 
        
        \draw[blue, thick] (N) -- (A);
        
        \draw[blue, thick] (P) -- (E);
        
        \draw[blue, thick] (I) -- (G);
        
        \draw[blue, thick] (K) -- (D);
        
        \draw[gray, thick] (A) -- (B) -- (C) -- (D) -- (E) -- (F) -- (G) -- (A);
        
        \draw[violet, thick] (A) -- (G);
        
        \draw[red, thick] (G) -- (F) -- (E);
        
        \draw[gray, thick] (H) -- (I) -- (J) -- (K) -- (L) -- (H);
        
        \draw[gray, thick] (M) -- (N) -- (O) -- (P) -- (M);
        
        \node at (0, 0) {$P$};

    \end{tikzpicture}
    \end{adjustbox}

\end{figure}

For arcs on an input polygon $P$, we only need to consider the arcs between adjacent endpoints of common outer tangents on $P$, since arcs between non-adjacent endpoints can be expressed as a union of these arcs. For example, in figure \ref{fig:cot}, the endpoints of common outer tangents associated with $P$ are $A$, $B$, $C$ and $D$. We only need to store the arc between $A$ and $B$, $B$ and $C$, $C$ and $D$, and $D$ back to $A$ (in clockwise direction). The arc from $A$ to $C$, for example, is the union of the arc from $A$ to $B$ and the arc from $B$ to $C$. Therefore, the number of arcs we need to consider is the same as the number of endpoints, which is also $O(n ^ 2)$. 

Techniques that computes the common outer tangents of two convex hulls in logarithmic time have been known since 1995 by Kirkpatrick et al.\cite{kirkpatrick}. Thus all $O(n ^ 2)$ common outer tangents can be computed in $O(n ^ 2 \log{m})$ time. The transitions for the dynamic programming process only involve common outer tangents and arcs between adjacent endpoints on imput polygons. Therefore, the dynamic programming process takes $O(n ^ 2)$ time per call and the total time is $O(n ^ 3)$.

However, we do need to maintain two look-up tables $g$ and $t$, where entries consist of common outer tangents and arcs. For table $g$ that stores values of $\Gamma$, entries for arcs are the harder part. Computation of the intersection between a ray and a convex polygon with at most $m$ vertices can be done by binary search based methods, which runs in $O(\log {m})$ time. This can be easily extended to intersection between the ray and an arc of the convex polygon. $g$ can therefore be updated $O(\log {m})$ per entry. Since there are $O(n)$ updates in total and there are $O(n ^ 2)$ entries, the total time complexity for maintaining table $g$ is $O(n ^ 3 \log {m})$

For table $t$ that checks if a tangent or an arc is intersection-free, consider the two types of updates: adding a polygon from $C$ to $C ^ {\prime}$, or merging a subset of $C ^ {\prime}$. For the first type, since the input polygons are disjoint, we only consider entries for tangents. Updates for these entries require only intersection testing between a line segments and a convex polygon with at most $m$ vertices, which can be done in $O(\log {m})$ time. The total time complexity is therefore also $O(n ^ 3 \log {m})$. For the second type, entries for arcs are the harder part. To check if an arc intersects the interior of a convex hull, we check if the endpoints of the arc is inside the convex hull and if the arc intersects the boundary of the convex hull at non-endpoints. Similar to the analysis for the main algorithm for this part, for a convex hull we get after merging subset $S$, we only need to go through $|S|$ edges and we check if the edge intersects a ray or the arc of a convex polygon with at most $m$ vertices. The former can be done in $O(1)$ time while the latter can be done in $O(\log{m})$ time. Therefore, the time complexity is $O(\log{m})$ per entry per edge. Since the number of edges involved in total is no more than $2n - 1$, we have a time complexity of $O(n \log {m})$ per entry and $O(n ^ 3\log {m})$ in total for this part. The total time complexity for maintaining $t$ is therefore $O(n ^ 3 \log {m})$.

Pre-processing takes $O(N)$ time. The total time complexity is therefore $O(n ^ 3 \log {m} + N)$, as promised.

\bibliographystyle{unsrt}
\bibliography{main}

\appendix

\begin{proof} [Proof of Theorem \ref{theo:submodular}]
    We first show that, if $A$ and $B$ are convex sets such that $A \subset B$, then for any point $p$, $\mathcal{H}(B \cup p) - \mathcal{H}(B) \le \mathcal{H}(A \cup p) - \mathcal{H}(A)$
    
    If $p \in B$, then $\mathcal{H}(B \cup p) - \mathcal{H}(B) = 0$, and $\mathcal{H}(A \cup p) - \mathcal{H}(A) \ge 0$.
    
    Suppose $p \notin B$. As shown in figure \ref{proofndlemma}, there exists points $B_u$ and $B_v$ such that the boundary of the convex hull of $B \cup p$ consists of line segments from $p$ to $B_u$ and $B_v$, and an arc of $B$ in the clockwise direction from $B_v$ to $B_u$, and there exists points $A_u$ and $A_v$ for $A$ similarly.
    
    Let $x$ be the intersection point between $p A_u$ and the convex hull of $B$, and $y$ be the intersection point between $A_v p$ and the convex hull of $B$. For any convex polygon $P$ and points $u, v \in P$, We use the notation $\overline{u v}_P$ to denote an arc of $P$ from $u$ to $v$ in the clockwise order.
    
    Consider the arc $\overline{A_u A_v}_A$ and the arc $\overline{B_u B_v}_B$ (both in blue). We have $\|\overline{A_u A_v}_A\| \le \|A_u x\| + \|\overline{x y}_B\| + \|y A_v\|$, $\|\overline{B_u B_v}_B\| \ge \|B_u x\| + \|\overline{x y}_B\| + \|y B_v\|$. From triangle inequality, we also have $\|B_u p\| - \|B_u x\| \le \|x p\|$ and $\|p B_v\| - \|y B_v\| \le \|p y\|$. Therefore:
    
    \begin{align*}
        \mathcal{H}(B \cup p) -  \mathcal{H}(B) 
        &= \|p B_v\|| + \|\overline{B_v B_u}_B\| + \|B_u p\| - (\|\overline{B_u B_v}_B\| + \|\overline{B_v B_u}_B\|) \\
        &= \|p B_v\| + \|B_u p\| - \|\overline{B_u B_v}_B\| \\
        &\le \|p B_v\| + \|B_u p\| - (\|B_u x\| + \|\overline{x y}_B\| + \|y B_v\|) \\
        &\le (\|B_u p\| - \|B_u x\|) + (\|p B_v\| - \|y B_v\|)  - \|\overline{x y}_B\| \\
        &\le \|x p\| + \|p y\|  - \|\overline{x y}_B\| \\
        &\le (\|A_u x\| + \|x p\|) + (\|p y\| + \|y A_v\|) - (\|A_u x\| + \|\overline{x y}_B\| + \|y A_v\|) \\
        &\le \|A_u p\| + \|p A_v\| - \|\overline{A_u A_v}_A\| \\
        &\le (\|A_u p\| + \|p A_v\| + \|\overline{A_v A_u}_A\|) - (\|\overline{A_u A_v}_A\| + \|\overline{A_v A_v}_u\|) \\
        &\le \mathcal{H}(A \cup p) -  \mathcal{H}(A) 
    \end{align*}
    
    To extend this to the case where where instead of a point $p$ we have another convex polygon $x$, it suffices to add the vertices of $x$ one by one to $A$ and $B$, and use induction.
\end{proof}

\begin{figure}[ht]

    \caption{} \label{proofndlemma}

    \centering

    \begin{adjustbox}{max width=1\textwidth}
    \begin{tikzpicture}[use Hobby shortcut]
        
        \coordinate (z0) at (-2.25, 0.46); 
        \coordinate (z1) at (-1, 1); 
        \coordinate (z2) at (1.53, 1);
        \coordinate (z3) at (2.39, 0.54);
        \coordinate (z4) at (-0.25, -1.16); 
        
        \draw[closed] (z0) .. (z1) .. (z2) .. (z3) .. (z4);
        
        \begin{scope}
            \clip (-2.25, 0.54) rectangle (4, 4);
            \draw[blue, closed] (z0) .. (z1) .. (z2) .. (z3) .. (z4);
        \end{scope}
        
        \begin{scope}
            \clip (0, 0.46) rectangle (4, 4);
            \draw[blue, closed] (z0) .. (z1) .. (z2) .. (z3) .. (z4);
        \end{scope}
                
        \coordinate (z0) at (-3, 1); 
        \coordinate (z1) at (-1.25, 1.68); 
        \coordinate (z2) at (2.03, 1.68);
        \coordinate (z3) at (3.69, 0.62);
        \coordinate (z4) at (1.65, -1.68);
        \coordinate (z5) at (-0.99, -2.26);
        \coordinate (z6) at (-2.91, -0.72); 
        
        \draw[closed] (z0) .. (z1) .. (z2) .. (z3) .. (z4) .. (z5) .. (z6);
         
        \begin{scope}
            \clip (-2.85, 1.2) rectangle (4, 4);
            \draw[blue, closed] (z0) .. (z1) .. (z2) .. (z3) .. (z4) .. (z5) .. (z6);
        \end{scope}
        
        \begin{scope}
            \clip (0, 1.18) rectangle (4, 4);
            \draw[blue, closed] (z0) .. (z1) .. (z2) .. (z3) .. (z4) .. (z5) .. (z6);
        \end{scope}
        
        \filldraw[black] (-0.19,3.48) circle (2pt) node[anchor=west] {$p$};
        
        \draw[thick] (-2.85, 1.18) -- (-0.19, 3.48);
        
        \draw[thick] (3.3, 1.20) -- (-0.19, 3.48);
        
        \draw[thick] (-2.25, 0.46) -- (-0.19, 3.48);
        
        \draw[thick] (2.39, 0.54) -- (-0.19, 3.48);
        
        \filldraw[black] (-2.85,1.18) circle (2pt) node[anchor=west] {$B_u$};
        
        \filldraw[black] (3.3,1.20) circle (2pt) node[anchor=west] {$B_v$};
        
        \filldraw[black] (-2.25, 0.46) circle (2pt) node[anchor=west] {$A_u$};
        
        \filldraw[black] (2.39, 0.54) circle (2pt) node[anchor=west] {$A_v$};
        
        \filldraw[black] (-1.44, 1.65) circle (2pt) node[anchor=west] {$x$};
        
        \filldraw[black] (1.31, 1.74) circle (2pt) node[anchor=west] {$y$};
        
        \draw (-2.85,1.18) -- (-1.44, 1.65);
        
        \draw (1.31, 1.74) -- (3.3, 1.20);
        
%        \begin{scope}
%            \clip (2.6, 5.36) rectangle (10, 10);
%            \draw[red, closed] (z0) .. (z1) .. (z2)
%        \end{scope}
        
%        \draw[red] (1.75, 2.3) -- (4.3, 5.36);
        
%        \node[text width=5cm] at (4.5, 4) {\large $A$};

        \node[text width=5cm] at (2.5, 0) {\large $A$};
        
        \node[text width=5cm] at (2.6, 1.5) {\large $B$};

    \end{tikzpicture}
    \end{adjustbox}

\end{figure}

\begin{proof} [Proof of Theorem \ref{theo:indivisible}]
    Suppose $C = O_1 \cup O_2 \cdots O_L$ is an optimal partition.

    Consider $\{O_i \cap {S} \mid O_i \cap {S} \ne \emptyset\}$. It's easy to see that this defines a partition of $S$. Since ${S}$ is indivisible, the answer we get from this partition is no less (therefore no better than) than the perimeter of the convex hull of $S$. 
    
    We define two partitions of $S$, $X$ and $Y$, in the following way: Initially, let $X$ be the trivial partition defined by $S = X_1$ and $Y$ be the partition defined by $\{O_i \cap {S} \mid O_i \cap {S} \ne \emptyset\}$. Initially the answer for $X$ is no less than the answer for $Y$, and $X_1 = S$. Our idea is that, we gradually add elements of $C \backslash S$ to $X$ and $Y$ such that $X$ and $Y$ start as partitions of $S$ and end up as partitions of $C$, while at any moment, the answer for $Y$ remains no less than the answer for $X$, and $S \subset X_1$. At the end of the process, $Y$ is equal to the optimal partition $O$. Since the answer for $X$ is no worse than $Y$, $X$ is also optimal. Since $S \subset X_1$, the proof is done. 
    
    For each $i$, if $O_i \cap S = \emptyset$, add $O_i$ as a new component to both $X$ and $Y$. If $O_i \cap S \ne \emptyset$, $O_i \cap S$ must be contained in a component $Y_k$ of $Y$. We add every $x \in O_i \backslash (O_i \cap S)$ one by one to $X_1$ for $X$, and to $Y_k$ for $Y$. It is easy to see that $Y_k \subset X_1$ at any moment of the adding process, and therefore from Theorem \ref{theo:submodular}, the answer for $X$ remains no less than the answer for $Y$. we therefore finish the proof.
\end{proof}

%\bibliography{references}  %%% Remove comment to use the external .bib file (using bibtex).
%%% and comment out the ``thebibliography'' section.

%%% Comment out this section when you \bibliography{references} is enabled.

\end{document}